# The role of carbon segregation on nanocrystallisation of pearlitic steels processed by severe plastic deformation.

X. Sauvage[1*] and Y. Ivanisenko[2]

1- Groupe de Physique des Matériaux - UMR CNRS 6634, Institut of Material Research, Université de Rouen, 76801 Saint-Etienne-du-Rouvray, France.

2- Institut für Nanotechnologie, Forschungszentrum Karlsruhe, 76021 Karlsruhe, Germany

*corresponding author : Xavier Sauvage
xavier.sauvage@univ-rouen.fr
Tel : + 33 2 32 95 51 42
Fax : + 33 2 32 95 50 32

**abstract**
The nanostructure and the carbon distribution in a pearlitic steel processed by torsion under high pressure was investigated by three-dimensional atom probe. In the early stage of deformation (shear strain of 62), off-stoichiometry cementite was analysed close to interphase boundaries and a strong segregation of carbon atoms along dislocation cell boundaries was observed in the ferrite. At a shear strain of 300, only few nanoscaled off-stoichiometry cementite particles remain and a nanoscaled equiaxed grain structure with a grain size of about 20 nm was revealed. 3D-AP data clearly point out a strong segregation of carbon atoms along grain boundaries. The influence of this carbon atom segregation on the nanostructure formation is discussed and a scenario accounting for the nanocrystallisation during severe plastic deformation is proposed.

## 1. Introduction

Processing of bulk nanocrystalline materials thanks to severe plastic deformation (SPD) has been widely investigated in the past two decades. Several SPD techniques have been designed to optimise grain refinement like Equal Channel Angular Pressing (ECAP) [1], High Pressure Torsion (HPT) [1-10] or Accumulated Roll Bonding (ARB) [11]. In pure metals, grain sizes in a range of 100 to 500 nm are commonly achieved. Such grain size reduction is usually attributed to the formation of dislocation cell walls which progressively become low angle grain boundaries and finally through higher level of deformation turn into high angle grain boundaries [5, 6]. It has been shown however that in severely deformed pearlitic steels, the grain size could be one order of magnitude smaller than that in pure iron [12, 13]. Pearlitic steels exhibit a lamellar structure made of a mixture of ferrite and of cementite ($Fe_3C$ carbide, volume fraction of about 12 %) [14] and it is now well admitted that cementite could be at least partly decomposed during plastic deformation [13, 15-20]. Previously published data show indeed that cementite is dissolved in pearlitic steels processed by HPT and since X-ray diffraction data did not show any significant $\alpha$-Fe peaks shift, it is though that carbon atoms have segregated along grain boundaries and dislocations [13]. Such carbon atom segregation along grain boundaries was recently reported for ball milled pearlitic steel where the cementite phase was also decomposed [21]. This feature could explain why the grain size reduction is more pronounced in such carbon steel than in pure iron.

The nanostructure formation of pearlitic steels subjected to SPD have been extensively studied by TEM, but quantitative measurements of the carbon redistribution following the strain induced decomposition of cementite can only be obtained by a three-dimensional atom probe (3D-AP). The aim of this work was therefore to carry out some experiments using this technique to map out the carbon distribution in 3D for a better understanding of the role of diffusing alloying elements on the nanostructure formation during SPD.

## 2. Experimental

The investigated material is a carbon steel UIC 860 : 0.6-0.8 wt. % C, 0.8-1.3 wt. % Mn, 0.1-0.5 wt. % Si, 0.04 wt. % P (max), 0.04 wt. % S (max), Fe-balance. This steel was austenitised at 1223 K during 30 minutes and cooled in air to achieve a fully pearlitic microstructure with the thickness of ferrite and cementite lamellae of 210 and 40 nm, respectively. Then, it was deformed by HPT under a pressure of 7 GPa with one and five turns of torsion (N) at a constant rate of 1 turn/min (see details in reference [13]).

Three-dimensional atom probe (3D-AP) and field ion microscopy (FIM) specimens were prepared by electropolishing [15]. Small rods were cut out from HPT discs and needle shaped specimens were prepared so that the tip was located at a distance of 3±0.5 mm from the disc centre (see reference [16] for details). The corresponding shear strain was $\gamma = 62$ ( $\pm 10$ ) and 300 ($\pm 50$) for N = 1 and 5, respectively [13]. FIM images were obtained at 80K, using Ne as imaging gas. 3D-AP analysis was carried out at 80K in UHV (residual pressure $10^{-8}$ Pa), with 20% pulse fraction and 1.7 kHz pulse repetition rate. The atom probe was equipped with a reflectron device to enhance the mass resolution and a CAMECA's position sensitive detector (Energy Compensated Optical Tomographic Atom Probe, ECoTAP). During 3D-AP analyses, carbon atoms are collected as ions ($C^+$, $C^{2+}$) and molecular ions ($C_3^{2+}$, $C_2^+$, $C_4^{2+}$, $C_3^+$) [22]. 3D-AP reconstructed volumes shown in this paper exhibit the distribution of all these different ions. Measurements of carbon concentration were performed following the procedure described by Sha and co-authors [22].

## 3. Results

Experimental data collected on the undeformed pearlitic steel are not shown in the present paper because they are consistent with already published data [15-18]. The measured carbon concentration is below 0.1 at% in the α-Fe phase and 25 ± 0.5 at.% in cementite lamellae ($Fe_3C$ carbides). Moreover, sharp gradients (less than 2 nm) were always detected across α-Fe / $Fe_3C$ interfaces.

*Carbon atom distribution after one turn of torsion under high pressure*

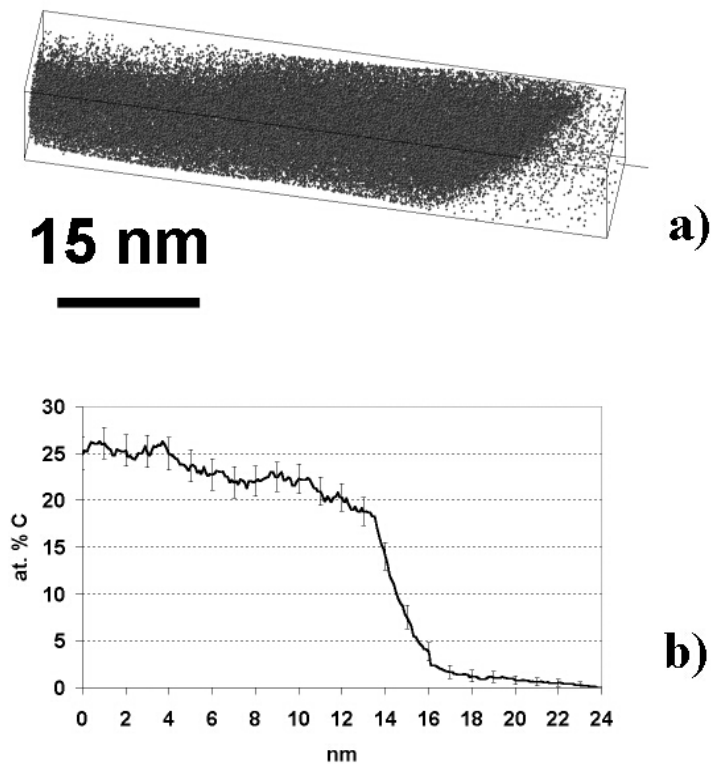

*Figure 1 : 3D reconstruction of an analysed volume in the pearlitic steel processed one turn by HPT (a). Only carbon atoms are plotted to exhibit a cementite lamellae. The carbon concentration profile was computed with a two nanometers sampling box across the ferrite/cementite interface (b).*

After one turn of torsion under high pressure, large regions of cementite were detected with the 3D-AP. Such a region is shown in the Fig. 1 (a) where only carbon atoms are exhibited. This is most probably a small part of a former cementite lamellae that might have been fragmented and plastically deformed. The composition profile computed across the α-Fe/cementite interface exhibits a sharp carbon gradient at this interface (two nanometers wide corresponding to the thickness of the sampling box). However, interesting features appear both on the $Fe_3C$ side and in the α-Fe side. In the far left of the profile, the carbon concentration is 25 at.% as expected for cementite, however along the interface there is a 10 nm thick layer with a carbon content in a range of 20 to 25 at.% which could be attributed to off-stoichiometry cementite. On the ferrite side, it is interesting to note as well that a large carbon gradient appears : there is about 2 at.% carbon in the ferrite close to the interface and this value slowly decreases down to zero at a distance of about 8 nm from the interface.

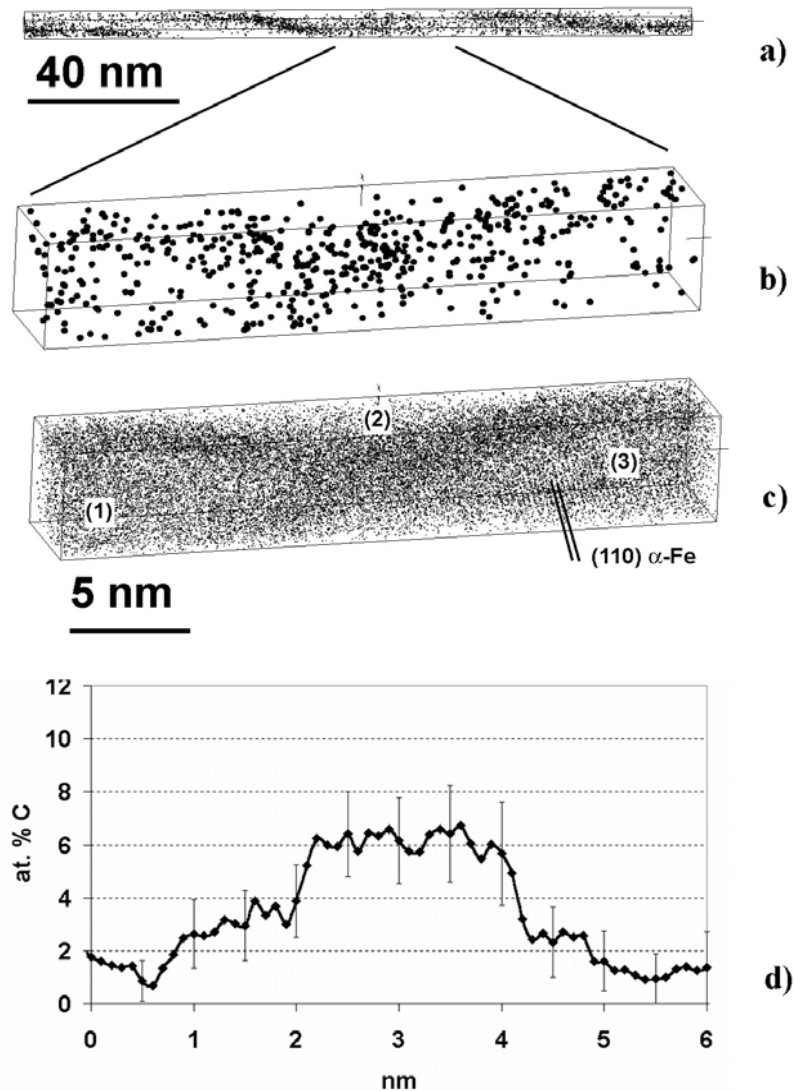

*Figure 2: Carbon atom distribution in an analysed volume in the pearlitic steel processed one turn by HPT (a). A small part of the volume was enlarged to show at higher magnification carbon (b) and iron distribution (c). Carbon atoms are segregated on planar defects separating ferrite zones labelled 1, 2 and 3. Atomic planes attributed to (110) α-Fe are clearly exhibited in zone 3. A composition profile was computed with a 1 nm sampling box across a carbon enriched planar defect (d).*

In the ferrite, thin bands containing a significant amount of carbon were also detected (Fig. 2 (a)). Their 3D shape is somewhat complicated : they are curled and twisted. As shown by the composition profile computed across one of them, their thickness is in a range of 2 to 3 nm and their carbon content is about 6 at.% (Fig. 2 (d)). Both the carbon and iron distribution is also shown at a higher magnification (Fig. 2 (b) and (c)) to point out three regions (labelled 1, 2 and 3) separated by carbon rich layers. The atomic resolution was achieved in the region 3 where (110) α-Fe atomic planes are clearly exhibited because they were almost perpendicular to the analysis direction. In the other two regions (labelled 1 and 2), these atomic planes are not exhibited. This indicates that there is a significant misorientation between these three regions and that they are three different ferrite grains. The local density in the reconstructed volume showing Fe atoms appears slightly higher at boundaries between ferrite grains (Fig. 2 (c)). This feature is attributed to local magnification effects resulting from a lower evaporation field of these boundary regions which contain a significant amount of carbon.

*Carbon atom distribution after five turns of torsion under high pressure*

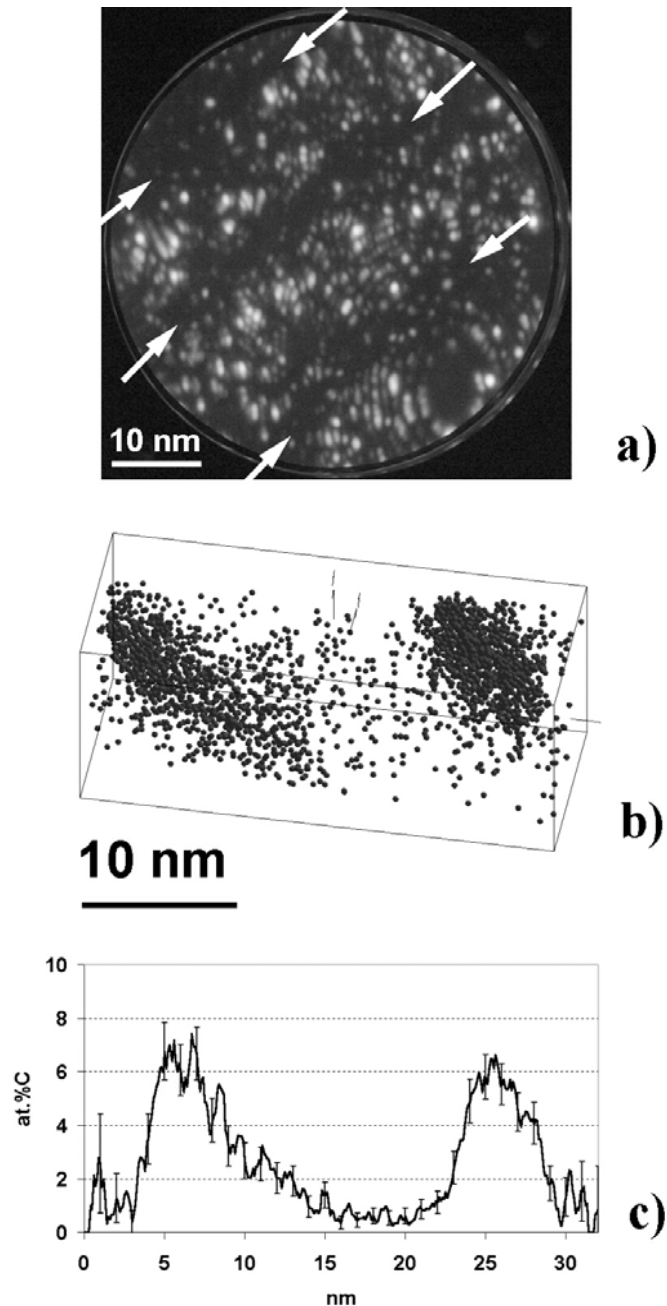

*Figure 3: Field ion microscopy image of the pearlitic steel processed five turns by HPT exhibiting a lamellar nanostructure (a). 3D reconstruction of an analysed volume where only carbon atoms are plotted to show two carbon rich lamellae (b). The composition profile was computed with a 2 nm sampling box across the two carbon rich lamellae (c).*

After five turns of torsion, two kinds of nanoscaled structures were observed by FIM within samples. Some regions exhibit a lamellar-like structure with an interlamellar spacing in a range of 10 to 20 nm as shown in the Fig. 3(a) where three dark lamellae are arrowed. Two of these lamellae were analysed with the 3D-AP, and the volume displayed in the Fig. 3(b) show that they contain a significant amount of carbon. Quantitative carbon concentration measurements were obtained thanks to a composition profile (Fig. 3(c)). They contain about 6 at.% carbon, and their thickness is about 5 nm.

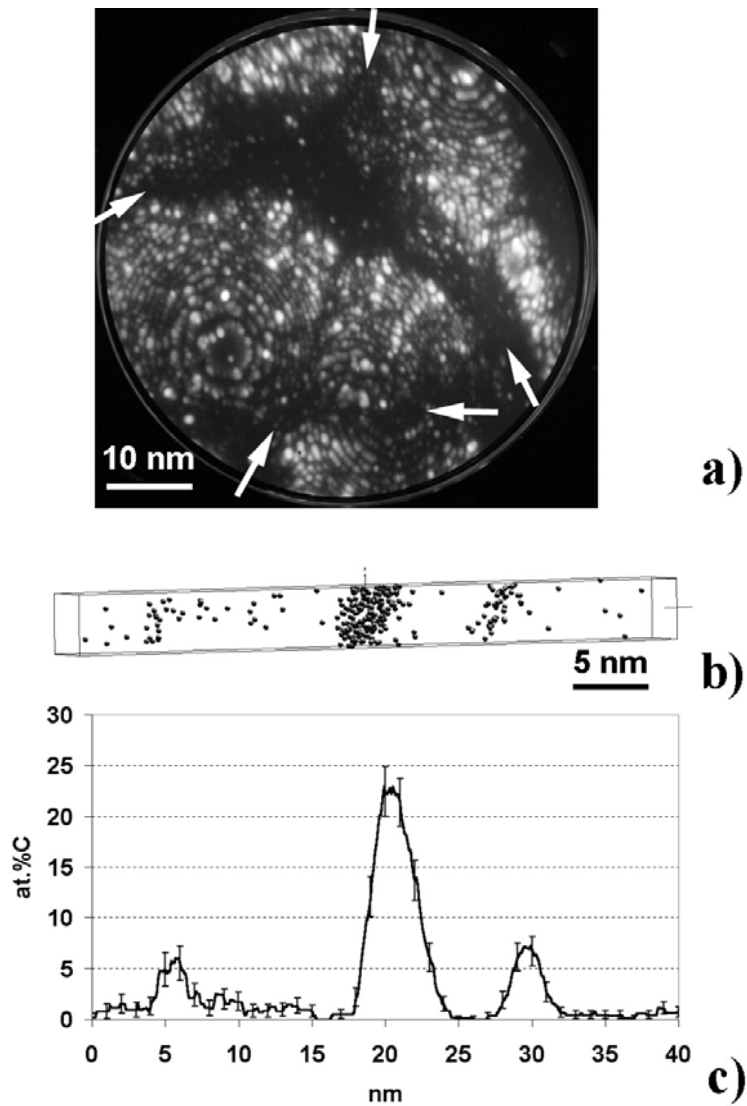

*Figure 4: Field ion microscopy image of the pearlitic steel processed five turns by HPT (a). Grain boundaries are arrowed. 3D distribution of carbon atoms within a small analysed volume (b) and carbon concentration profile computed across with a 2 nm sampling box (c).*

Field ion microscopy observations revealed other regions with a very different kind of nanostructure. Such a region is shown in the Fig. 4 (a) where several grain boundaries are arrowed. The grain size is in a range of 10 to 30 nm, and grains are equiaxed. The distribution of carbon atoms is shown by the small analysed volume displayed in the Fig. 4(b) and the composition profile computed through this volume (Fig. 4(c)). A nanoscaled cementite particle is located in the middle of the volume. As shown by the profile, cementite is slightly off-stoichiometry and contains between 20 and 25 at. %C. On the left and on the right of the particle, segregation of carbon atoms along two planar defects is clearly exhibited. They are most probably the grain boundaries observed on the FIM image (Fig. 4(a)).

## 4. Discussion

*Decomposition of cementite*
The 3D-AP analyses confirm that cementite is decomposed as it has been previously reported [13], but the data shown here provide new information about the decomposition mechanism. The cementite particle analysed after one turn of torsion clearly exhibits that its structure is not homogenous. It contains 25 at.% in the core, while in a thin boundary layer it is slightly off-stoichiometry (Fig. 1). Moreover, 3D-AP data clearly show that even after 5 turns, cementite decomposition is not completed (Fig. 4) but cementite particles are much larger at lower strains probably because they are less fragmented but also less dissolved. It is interesting to note that after five turns of torsion, $Fe_3C$ particles are so small that they are fully off-stoichiometry (Fig. 4). Thus, the decomposition of cementite during SPD may proceed in two steps : first, C-vacancies are introduced in the $Fe_3C$ phase which leads to the formation of a metastable off-stoichiometry cementite [23-25]. Then, in a second step this latter structure is fully decomposed. As argued by Gridnev and Gavrilyuk, dislocations in the ferrite lattice could play an important role in this process [19]. They proposed that the transfer of carbon atoms from cementite to dislocations in the ferrite is energetically favourable since the interaction energy between carbon atoms and dislocations is higher than the bonding energy between Fe and C atoms in the $Fe_3C$ phase (0.8 eV/atom and 0.5 eV/atom respectively). Thus, close to the interphase boundary, carbon atoms may jump from their lattice site in the cementite phase to dislocations located in the ferrite. This mechanism is consistent with the present experimental data showing a thin layer of off-stoichiometry cementite along the $Fe_3C/\alpha$-Fe interface (Fig.1(b)).

*Diffusion of carbon atoms through the ferrite*
After one turn of torsion, a significant carbon gradient was exhibited in the ferrite close to the $Fe_3C/\alpha$-Fe interface with a maximum concentration in a range of 1 to 2 at.% (Fig. 1(b)). Since the solubility of carbon in BCC iron is very low [14], the formation of such a super saturated solid solution is unlikely to happen. Moreover it has already been confirmed by X-ray diffraction measurements showing that the lattice parameter of the $\alpha$-Fe phase is not affected by $Fe_3C$ decomposition even after 5 turns of torsion [13]. Therefore, this observed gradient is not attributed to downhill diffusion of interstitial C atoms in the BCC ferrite lattice but to carbon atom segregation along dislocations segments pinned by $Fe_3C/\alpha$-Fe interfaces. In this case the homogenous distribution of carbon observed in small 3D-AP reconstructions (Fig. 1(b)) can be attributed to overlapping Cottrell atmospheres from neighbouring dislocations. Broad segregation of carbon atoms around dislocations (up to 7nm from the core) have indeed been reported by Wilde and co-authors [26]. Moreover, TEM observations of severely deformed pearlite always demonstrate the formation of very high dislocation density in the ferrite close to the inter-phase boundary [12,13]. If it is assumed that this density is about $\rho=10^{12}$ cm$^{-2}$ (typical density in severely deformed metals [1]), then the mean distance between dislocations is $L \sim 1/\sqrt{\rho} = 10$ nm. In such a configuration, Cottrell atmospheres from neighbouring dislocations could simply overlap, and a "homogenous" distribution will be observed by 3D-AP. Such a segregation of carbon atoms to dislocations is consistent with the mechanism proposed by Gridnev and Gavrilyuk [19].
On the concentration profile displayed in the Fig. 1(a), it is interesting to note that there are more carbon atoms missing in the $Fe_3C$ than carbon atoms detected in the ferrite. This indicates that some of them have diffused over long distances. Several mechanisms should be considered : bulk diffusion, dislocation drag and pipe diffusion. However, the deformation was performed at room temperature and no significant increase of the temperature is expected during the HPT process [13]. In such conditions, the mobility of interstitial carbon atoms is very low in the ferrite and as previously reported [13], the estimated diffusion distance of carbon is much lower than 1 nm (during the deformation time which is about 300 s). Obviously, from the Fick's law ( $J = -D\,\text{grad}(C)$ ), such a low diffusivity (D) combined with the low solubility (C) of carbon in the BCC ferrite [14] cannot give rise to a significant carbon flux (J) through bulk diffusion. Moreover, this very low mobility of carbon makes as well the dislocation drag mechanism not realistic. Thus, carbon atoms might diffuse through pipe diffusion in dislocation cores. It is indeed well known that bulk diffusion rates of solute elements is much lower than their diffusion rates along dislocation cores where the activation energy could be lowered by a factor of two [27].

*Transformation of the pearlitic structure during HPT*
The transmission electron microscopy investigation of the microstructure evolution of a pearlitic steel of the same composition during torsion under high pressure had been carried out in [13] and general trends can be summarized as follows. At the beginning of HPT deformation (e.g. after one turn of torsion), a very high dislocation density is observed in the ferrite phase in the vicinity of ferrite / cementite interphase boundary. Then, between one and three turns of torsion, a cellular structure forms in the ferrite phase, pearlite colonies progressively align along the shear direction and cementite platelets are elongated. Finally, higher shear strains lead to a gradual increase of misorientations of cell boundaries in the ferrite and to their transformation into high angle grain boundaries. This process is accomplished after five turns of rotation and it gives rise to a very homogenous nanocrystalline structure with a grain size in the range of 10 to 20 nm where cementite is almost completely decomposed [13, 28]. This evolution of the microstructure of pearlitic steels during HPT is consistent with the observations reported for other processes like wire drawing [17, 18, 29], ball milling [21, 30], shot peening [31] and on surfaces of railway rails [12, 32].

The present data are consistent with TEM observations described above but they reveal quite new features. After five turns of torsion (N=5), the original lamellar structure of the pearlite is strongly affected by the plastic deformation. Two kinds of nanoscaled structure were revealed by the three-dimensional atom probe analyses. First, some regions still exhibit a lamellar structure which is very similar to the microstructure of pearlitic steels processed by cold drawing (steel cords) [17,18]. Carbon rich lamellae containing less than 10at.% carbon are exhibited, the interlamellar spacing is smaller than 20 nm and large carbon gradients appear (Fig. 3). Such carbon rich lamellae are former $Fe_3C$ lamellae that were strongly elongated and partly decomposed during the HPT process. However, only few regions with this kind of lamellar structure were revealed (less than 20% vol.). Other colonies have been completely transformed into an equiaxed structure made of nano-scaled $\alpha$-Fe grains stabilised by carbon atoms segregated along grain boundaries (Fig. 4). Such a microstructure is similar to the microstructure obtained by Ohsaki and co-authors by mechanical milling of a pearlitic steel [21]. During mechanical milling, impacts and friction of steel balls induce the severe plastic deformation. This process is not continuous and the strain rate is much higher than during HPT. This could explain why a smaller grain size was achieved (about 10 nm) and why a significant amount of carbon was detected in solid solution in the ferrite (up to about 1at.%). However, it is interesting to note that similar carbon concentrations along grain boundaries were reported (close to 6at.%).

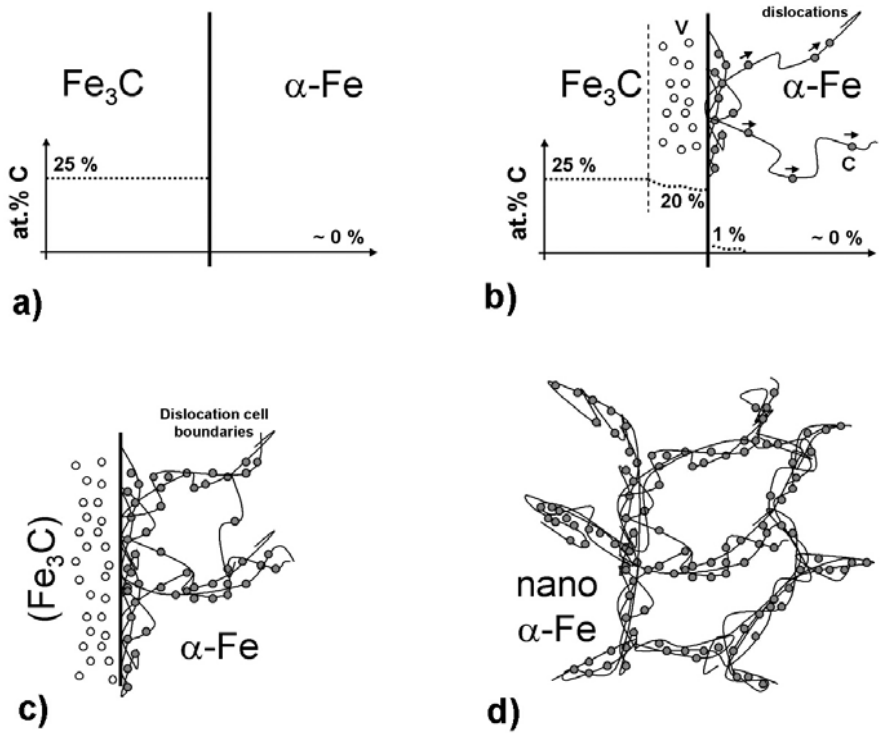

*Figure 5: Schematic representation of the nanostructure formation in the pearlitic steel processed by HPT. Original microstructure made of stoechiometric Fe₃C and α-Fe phases (a) ; Jump of C atoms (grey dots) from Fe₃C to dislocations in the ferrite and formation of C-vacancies (open circles) in the Fe₃C (b) ; Saturation of cell boundaries with C atoms, increase of misorientation between cells, further decomposition of Fe₃C and formation of new cell boundaries (c) ; Final microstructure made of α-Fe nanograins stabilised by carbon segregation along grain boundaries (d).*

On the base of previous TEM observations [13] supported by the present 3D AP data we suggest the following mechanism of the formation of the equiaxed nanoscaled structure during HPT processing of the pearlitic steel, sketched out in the Fig. 5. The original microstructure (N=0) is made of α-Fe with an extremely small amount of carbon in solid solution [14] and stoichiometric Fe₃C lamellae containing 25at.% carbon Fig. 5(a). At the early stage of deformation (N <1), Fe₃C lamellae are elongated and fragmented. Carbon atoms in Fe₃C nearby interphase boundaries jump to dislocations in the ferrite leaving vacancies and a thin layer of off-stoichiometry Fe₃C (data Fig. 1 and sketch Fig. 5(b)). Then, some of them diffuse along the core of dislocations collected at Fe₃C/α-Fe interfaces. With increasing of strains, dislocation cells develop in the ferrite and cell boundaries trap carbon atoms (data Fig. 2 and sketch Fig. 5(c)). At higher level of deformation ($1 \leq N \leq 5$), Fe₃C lamellae are further fragmented and decomposed via the mechanism discussed above. Dislocations previously collected at Fe₃C/α-Fe boundaries are saturated with carbon atoms and are locked in the ferrite. Dislocations generated in the course of deformation are frequently pinned with carbon atoms coming from cementite decomposition and that diffuse in ferrite lamellae along dislocation cores. These pinned dislocations initiate the formation of new cell boundaries situated at rather small distances from each other. Simultaneously, an increase of the misorientation between dislocation cells occurs due to sink of not-pinned dislocations in cell boundaries. That way, a smaller cell size than in pure iron after similar level of deformation is obtained [5, 6]. Cementite particles are progressively decomposed and after five turns of torsion (N=5), only few Fe₃C nanoclusters remain (data Fig. 4). Most of carbon atoms are located along former dislocation cell boundaries that have progressively become new grain boundaries stabilised by carbon atoms (data Fig. 4 and sketch Fig. 5(d)).

## 5. Conclusions

i) After one turn of torsion under high pressure (shear strain of 62), off-stoichiometry cementite with a carbon concentration in a range of 20 to 25 at.% was analysed in the vicinity of interphase boundaries. Carbon atoms released by this decomposition of $Fe_3C$ were detected in the ferrite close to the interphase boundary or along dislocation cell boundaries.
ii) After five turns of torsion under high pressure (shear strain of 300), only few nanoscaled off-stoichiometry cementite particles remain. The microstructure is mostly equiaxed with a grain size of about 20nm but few regions with a lamellar structure still exist. Carbon atoms are segregated along grain boundaries with a typical carbon concentration peak of about 6 at.%.
iii) These experimental data confirm the mechanism proposed by Gridnev and Gavrilyuk involving dislocations [19] : during the plastic deformation, C-vacancies are created in cementite and carbon atoms jump in the core of dislocations located in the ferrite close to interphase boundaries.
iv) It is proposed that carbon atoms diffuse along dislocation cores through pipe diffusion. Dislocations are progressively pinned and serve as origin of new dislocation boundaries. This leads to a smaller cell size than in pure iron. Further deformation induces higher decomposition rate of cementite, saturation of dislocation cell boundaries with carbon atoms and increasing of misorientations. This mechanism is though to promote the nanocrystallisation of pearlite during HPT.

**Acknowledgements**
Professor R.Z. Valiev (Institute of Physics of Advanced Materials, Ufa, Russia) is gratefully acknowledged for the processing of the samples by torsion under high pressure.